\documentclass[twocolumn,rmp,amsmath,amssymb,showpacs]{revtex4}

\usepackage{graphicx}
\usepackage{dcolumn}
\usepackage{bm}

\begin{document}

 \title{Effect of dye doping on the charge carrier balance in PPV light emitting diodes as measured by admittance spectroscopy}

 \author{I. N. Hulea}\altaffiliation{also at the Dutch Polymer Institute, Eindhoven}

 \author{R.F.J. van der Scheer}\altaffiliation{also at the Eindhoven University of Technology}

 \author{H. B. Brom}

 \affiliation{Kamerlingh Onnes Laboratory, Leiden University, P.O. Box 9504, 2300 RA Leiden, The Netherlands}

 \author{Bea M. W. Langeveld-Voss}
 \affiliation{TNO Industrial Technology, De Wielen 6, 5600 HE Eindhoven,  The Netherlands}

 \author{A. van Dijken}

 \author{K. Brunner}
 \affiliation{Philips Research Laboratories, Prof. Holstlaan 4, 5656 AA Eindhoven, The Netherlands}

 \date{\today}

\begin{abstract}
Dye doping is a promising way to increase the spectral purity of
polymer light-emitting diodes (LEDs). Here we analyze the frequency
and field dependence of the complex admittance of
Al-Ba-PPV-PEDOT-ITO LEDs with and without dye. We compare the charge
carrier mobilities of pristine and dye-doped double-carrier and
hole-only (Au replacing Al-Ba) devices. Dye doping is shown to
significantly influence the electron mobilities while the hole
mobilities are left unchanged and thereby changing the carrier
balance in a double carrier device towards that of a hole only device. The minimum
in the LED capacitance as function of voltage appears to be an
excellent probe for the electron trapping phenomenon underlying the
reduction of the mobility.
\end{abstract}


\maketitle

\begin{figure}[h,t,b]
\begin{center}
 \includegraphics[width=8cm]{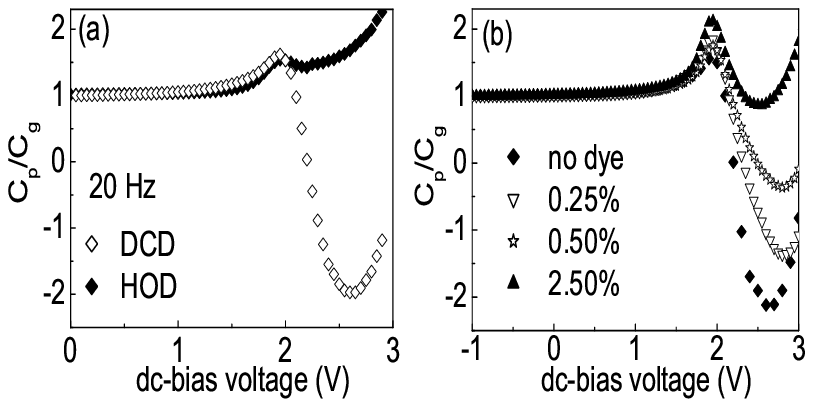}
\end{center}
\noindent{\caption{(a) $C_p/C_g$ versus bias voltage for a HO and DC
device (HOD resp. DCD) at 20~Hz. The HOD curve is shifted by 1.50~V
towards more positive bias to overlap $V_{bi}$. (b) $C_p/C_g$ versus
bias voltage for different dye-doping concentrations in a DCD. }
\label{Cp}}
\end{figure}

\begin{figure}[t]
\begin{center}
 \includegraphics[width=7cm]{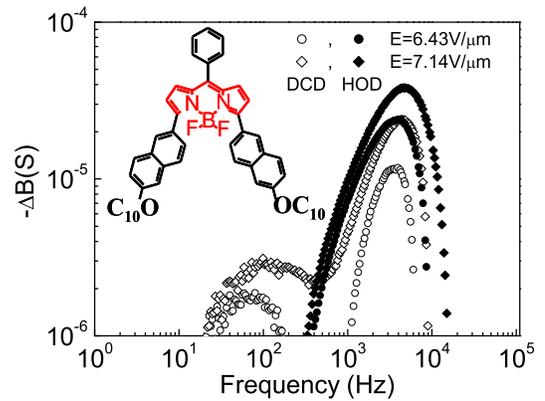}
\end{center}
\noindent{\caption{Determination of $\mu_e$ and $\mu_h$ from
$-\Delta B=-\omega(C-C_g)$ versus $\omega/2\pi$ for the $2.5~\%$
dye-doped sample. Inset shows the molecular structure of the dye.}
\label{mu}}
\end{figure}

Although an obvious route for a true full-color light emitting
diodes (LEDs) is employing three different polymers for the primary
colors, there are several other approaches currently being examined.
A particular appealing solution is energy or charge transfer from a
host polymer to emissive guest molecules. This process can be
optimized to virtually hundred percent efficiency already at low
guest concentrations, which should leave the macroscopic properties
of the polymer, such as viscosity, unchanged. The latter is
particularly important for the efficient production of pixelated
devices by ink-jet printing because the printing process has to be
optimized only for one ink instead of three. For these applications
detailed knowledge about the changes of the charge carrier
properties upon adding guest molecules to a host polymer is desired.
Low-frequency admittance data of LEDs has been proven to give
valuable information about the mobility of charge carriers
\cite{Martens00}. Here we measure the diode admittance as a function
of bias voltage at low frequencies and compare devices with and
without an emissive guest (dye dopant) in the active polymer layer.
When we increase the bias voltage $V$ above the built-in voltage
$V_{bi}$, the diode capacitance $C_p$ shows a minimum that i) is
markedly different for hole-only (HO) and double-carrier (DC)
devices and ii) is strongly dependent on the doping concentration.
In pristine samples and at low doping concentrations a formally
negative $C_p$ is measured at low frequencies \cite{noteCneg}. By
comparing admittance data for DC and HO devices, the capacitance
minimum at a voltage just above $V_{bi}$ proves to be an excellent
probe for electron trapping and mobility. It will be shown that on
the time scale of our measurement dye doping in these devices
effectively changes the device from a double-carrier towards an
electron deficient device.\\
\indent All samples were prepared in a glove box under N$_2$
atmosphere. A polymer/toluene solution containing 4.5 mg/ml polymer
is stirred at room temperature over night and heated at 70$^{\circ}$
C for 1 hour before the dye is added from a stock solution in
toluene (4 mg/ml). After stirring again for 1 hour, the mixture is
filtered over a 5 $\mu$m filter. The glass substrates, covered with
a 120 nm transparent and patterned ITO (indium tin oxide) layer (
$<$ 20 $\Omega/\Box$), were exposed to ultraviolet radiation in an
ozone atmosphere for 10 min. prior to spin coating. Remaining dust
particles were blown away with ionized nitrogen before 150$\pm$5 nm
of a conducting polymer (poly(3,4-ethylenedioxythiophene:polystyrene
sulphonic acid, PEDOT:PSS, from Bayer AG) and 70$\pm$5 nm of an
emissive polymer were spin coated on the substrates. The diode
structures consist of Al-Ba-PPV-PEDOT:PSS-ITO for the DC device and
Au-PPV-PEDOT:PSS-ITO for the HO device. All devices have a surface
area of 100 mm$^{2}$ typically and are encapsulated. The yellow
emitting PPV-based polymer was prepared via the Gilch polymerization
route \cite{Becker00}. The synthesis of the red dye will be
published elsewhere and its structure can be found in Fig.~\ref{mu}.
From ultraviolet photoemission spectroscopy (UPS) (HOMO of the
polymer) and cyclo-voltammetry measurements (HOMO and LUMO of dye
and polymer) the lowest unoccupied and highest occupied molecular
orbitals (LUMO-HOMO) levels have been determined for both host and
dye. For the yellow PPV the HOMO is at 5.2 eV below the vacuum level
and the LUMO at 2.9 eV. For the dye, the HOMO is at 5.4 eV and the
LUMO at 3.5 eV. The levels for PEDOT-PSS and Al-Ba are at 5 eV and
2.3 - 2.6 eV, respectively.\\
\indent In impedance measurements a LED can be seen as a voltage and
frequency dependent capacitance ($C_p(\omega,V)$) and resistance
($R_p(\omega,V)$) in parallel, so that the admittance ($Y$) can be
written as the sum of the conductance $G=1/R_p$ and the susceptance
$B=\omega C_p$: $Y =G+iB$. $Y$ was measured with an Agilent 4284a
RCL meter. The amplitude of the ac voltage applied on top of the dc
bias (-2~V to 5~V) was 50 mV.\\
\indent Fig.~\ref{Cp}a shows $C_p/C_g$ as a function of dc bias for
a DC and HO device without dye at 20 Hz. The geometrical capacitance
$C_g$ is measured in the fully depleted state of the device at
negative voltages. Especially the region just above $V_{bi}$ will be
shown to contain valuable information about the changes in the
device as a function of dye doping. For a DC device $C_p(V)/C_g$
decreases very rapidly to a minimum of $-2$, while $R_p$ decreases
continuously. When the dc bias is increased further, the capacitance
increases again. The increase in capacitance below $V_{bi}$ has been
observed before for thermally converted precursor PPV \cite{Meier97}
and will be dealt with in a separate contribution. $C_p(V)$ and
$R_p(V)$ of a HO device can be described analogous to a DC device,
see Fig.~\ref{Cp}a. Due to the differences in the built-in voltages
the position of the maximum of $C_p(V)/C_g$ is at $0.45 \pm 0.05$~V
for the HO device and at $1.95 \pm 0.05$~V for the DC device. In
Fig.~\ref{Cp}a the HO curve has been shifted by 1.50~V so that its
maximum coincides with that of the DC curve. Both curves have a
(local) minimum in $C_p(V)/C_g$ above the built-in voltage $V_{bi}$.
In Fig.~\ref{Cp}b the effect of dye doping on the capacitance of a
DC device is shown. The minimum capacitance above the built-in
voltage gradually increases upon increasing the dopant concentration
and already for a concentration of 2.5~\% its value approaches that
of a HO device.\\
\indent Since the obvious difference between DC devices and HO
devices is the presence of electrons, it is tempting to associate
the changes of the capacitance minimum above the built-in voltage
upon dye doping to the response of the electrons participating in
the current transport. In order to substantiate this assumption the
charge transport properties of positive and negative charge carriers
have to be evaluated and brought into relation to $C_p$. Frequency
scans \cite{Martens00} provide a simple way to derive the mobility
($\mu$) by plotting $-\Delta B$ vs. $\omega/ 2 \pi$, see
Fig.~\ref{mu}. The characteristic time $\tau_r=1/\omega_r$ (at
$\omega_r$ the maximum in $\Delta B$ appears) is related to the
transit time $\tau_t$ of the charge carriers via $\tau_r = 0.29
\tau_t$ \cite{Martens00}. This nondestructive measurement gives
electron and hole mobilities at the same time and is done on the
device as used in applications.
If the electrical field ($E$) dependence of the carrier mobility
$\mu$ in these disordered organic semiconductors is described by
$\mu = \mu(0) \exp(\gamma \sqrt E)$, we obtain for the zero field
hole mobilities in the pristine HO {\it and} DC devices a value of
$\mu(0)_h=1.12 \pm 0.1\times10^{-11}{\rm m^2/Vs}$ with an activation
field factor $\gamma=7.12 \pm 0.7\times10^{-4}\,{\rm (m/V)^{1/2}}$.
The electron mobility in the pristine DC device is found to be
nearly two orders of magnitude lower $\mu(0)_e=5.7\pm
0.9\times10^{-13}\,{\rm m^2/Vs}$ with
$\gamma=8.9\pm 1.2\times10^{-4}\,{\rm (m/V)^{1/2}}$.\\
\indent For dye-doped samples $\mu_h$ hardly changes (within 10 \%)
with the dye concentration. This is not a surprising result since it
was established before by current-voltage scans of HO devices that
the dye does not influence the current density of such devices.
Furthermore, if the HOMO of a dopant is situated energetically below
the HOMO of the host, as is the case in our system, the hole
transport is not expected to be affected \cite{Malm02}. However, the
situation is considerably different for the electron mobilities. For
2.5~\% dye, $\mu(0)_e$ is reduced from $5.7\pm 0.9\times10^{-13}{\rm
m^2/Vs}$ to $3.8\pm 0.6\times10^{-14}{\rm m^2/Vs}$ and the
activation field factor is increased from $\gamma=8.9\pm
1.2\times10^{-4}{\rm (m/V)^{1/2}}$ to $\gamma=13.9\pm
2\times10^{-4}{\rm (m/V)^{1/2}}$. For a field of $7.5 \,{\rm V/\mu
m}$ $\mu_e$ will be $1.7\times10^{-12}\,{\rm m^2/Vs}$, about three
times lower than in the pristine sample.

\begin{figure}[t]
\begin{center}
 \includegraphics[width=7cm]{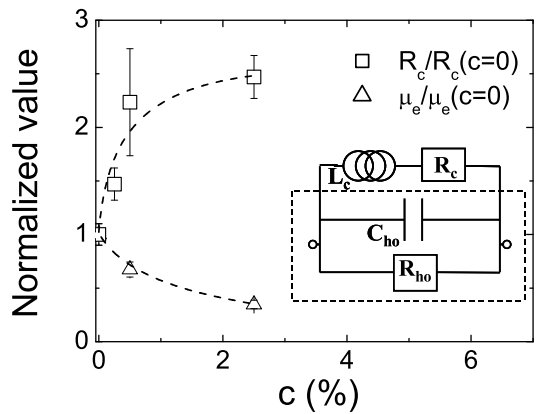}
\end{center}
\noindent{\caption{The dye concentration ($c$) dependence of $\mu_e$
and $R_c$ at $E = 7.5 \,{\rm V/\mu m}$ using the replacement circuit
(at 20~Hz) shown in the inset. All values are normalized to those in
the DC device without dye doping at $E= 7.5\, {\rm V/\mu m}$: $R_c=
1.7 {\rm k\Omega}$ and $\mu_e = 4.9 \times 10^{-12} {\rm m^2/VS}$.
$L_c \approx$~0.4~H depends hardly on $c$, but decreases with
increasing $E$ (for $c=0$ from 0.4~H at $E$ = 7.5 ${\rm V/\mu m}$ to
0.13~H at $E$ = 10.5 ${\rm V/\mu m}$, while $R_c$ decreases from 1.7
to 0.94~k$\Omega$. $C_{\rm ho}$ and $R_{\rm ho}$ are as measured for
the HO device: $C_{\rm ho} = 58.6$~nF and $R_{\rm ho} = 1.4 {\rm
k\Omega}$ for $E$ = 7.5 ${\rm V/\mu m}$}. \label{parameters}}
\end{figure}

The influence of the dye dopant on the total current of a DC device
is calculated at the fixed measurement frequency of 20 Hz using the
replacement circuit in Fig.~\ref{parameters}. The circuit includes
explicitly the experimental values determined for the HO device
(dashed box) to which in parallel the $R_c - L_c$ branch is added to
account for the electron current and the increase in hole current
compared to the HO device due to the changed charge distribution and
the recombination in a DC device. As mentioned above, the hole
current in the HO device is independent on dye doping. So $R_c$ and
$L_c$ take into account all the changes in the device related to the
presence of electrons. The inductance $L_c$ is needed to account for
the delay in the current in its response to the ac-field
\cite{Martens01}. Obviously, the combination of the unknown $R_c$
and $L_c$ together with the measured values of the HO have to give
the right total current in a DC at 20 Hz. This model is particularly
suited to show the {\it trends} upon dye doping and to present them
in a structured way.\\
\indent The dependence of $R_c$ and $\mu_e$ on $c$ is shown in
Fig.~\ref{parameters}, where the absolute values of these parameters
are given in the caption. The total current through the device is
mainly determined by the values of $R_{\rm ho}$ and $R_c$. In the DC
device without dye doping the current is about two times larger than
in the HO device (at voltages $<3$~V). Upon dye doping $R_c$
increases by a factor of three. The shallower minimum of the
capacitance in Fig.~\ref{mu} with dye doping is mainly caused by
this increase of the resistance $R_c$ in the $R_c-L_c$ branch. $R_c$
reduces the current through the branch and hence the relative
importance of $L_c$. The changes in the capacitance minimum
therefore are a sensitive probe for the changes in {\it extra}
conductance and are influenced by the introduction of the negative
charge carriers and the presence of dye dopants.\\
\indent When translating this macroscopic result into a microscopic
picture, in principle two borderline cases can be distinguished.
First, the extra current density might be predominantly due to
electrons. If so, the decrease in $\mu_e$ with increasing dye dopant
will already explain the increase in $R_c$. This possibility is not
likely, because in our devices $\mu_e << \mu_h$, see Fig.~\ref{mu}
and the number of electrons and holes are about equal \cite{Blom98}.
The most likely scenario is that the excess current is mainly
carried by holes (the other extreme). This is expected, when the
space charge clouds compensate each other and allow for additional
positive charge carriers to be injected. The region of this
compensation grows with increasing $\mu_e$ and hence the lower
$\mu_e$, the higher $R_c$, as observed in Fig.~\ref{parameters}. In
both scenario's the reduction of $\mu_e$ with dye doping can be seen
as a natural consequence of the electron trapping ability of the
molecular dopant and will govern the device properties of all
devices having a similar energy landscape as our model system.\\
\indent Summarizing, the data show that upon dye doping the DC
devices become electron deficient. The changes in $C_p$ turn out to
be very sensitive for this phenomenon and can be used as a probe for
the carrier balance and charge carrier trapping behavior in
polymeric
LEDs. \\
\indent We acknowledge discussions with Hans Huiberts (Philips),
Thijs Michels (TUE), Paul Blom (RUG), Herman Schoo and Jolanda
Bastiaansen (TNO), and Hubert Martens (Philips), who suggested the
connection between $\mu_e$ and the hole current, and thank Dr. Helga
Hummel (Philips) and Dr. Volker van Elsbergen (Philips) for the
cyclo-voltammetry resp. UPS measurements. The work forms part of the
research programme of the Dutch Polymer Institute (DPI).

\end{document}